\documentclass[preprint2]{aastex}

\shorttitle{Redshift/CO Modeling for HLSW-01}
\shortauthors{Scott et al.}

\begin{document}

%% LaTeX will automatically break titles if they run longer than
%% one line. However, you may use \\ to force a line break if
%% you desire.

%\title{Collapsed Cores in Globular Clusters, \\
%    Gauge-Boson Couplings, and AAS\TeX\ Examples}

\title{Redshift Determination and CO Line Excitation Modeling for the Multiply-Lensed Galaxy HLSW-01}

%% Use \author, \affil, and the \and command to format
%% author and affiliation information.
%% Note that \email has replaced the old \authoremail command
%% from AASTeX v4.0. You can use \email to mark an email address
%% anywhere in the paper, not just in the front matter.
%% As in the title, use \\ to force line breaks.

%\author{S. Djorgovski\altaffilmark{1,2,3} and Ivan R. King\altaffilmark{1}}
%\affil{Astronomy Department, University of California,
%    Berkeley, CA 94720}

%\author{C. D. Biemesderfer\altaffilmark{4,5}}
%\affil{National Optical Astronomy Observatories, Tucson, AZ 85719}
%\email{aastex-help@aas.org}

%\and

%\author{R. J. Hanisch\altaffilmark{5}}
%\affil{Space Telescope Science Institute, Baltimore, MD 21218}

\author{K.S.~Scott\altaffilmark{1},
R.E.~Lupu\altaffilmark{1},
J.E.~Aguirre\altaffilmark{1},
R.~Auld\altaffilmark{2},
H.~Aussel\altaffilmark{3},
A.J.~Baker\altaffilmark{4},
A.~Beelen\altaffilmark{5},
J.~Bock\altaffilmark{6,7},
C.M.~Bradford\altaffilmark{6,7},
D.~Brisbin\altaffilmark{8},
D.~Burgarella\altaffilmark{9},
J.M.~Carpenter\altaffilmark{6},
P.~Chanial\altaffilmark{3},
S.C.~Chapman\altaffilmark{10},
D.L.~Clements\altaffilmark{11},
A.~Conley\altaffilmark{12},
A.~Cooray\altaffilmark{6,13},
P.~Cox\altaffilmark{14},
C.D.~Dowell\altaffilmark{6,7},
S.~Eales\altaffilmark{2},
D.~Farrah\altaffilmark{15},
A.~Franceschini\altaffilmark{16},
D.T.~Frayer\altaffilmark{17},
R.~Gavazzi\altaffilmark{18},
J.~Glenn\altaffilmark{12,19},
M.~Griffin\altaffilmark{2},
A.I.~Harris\altaffilmark{20},
E.~Ibar\altaffilmark{21},
R.J.~Ivison\altaffilmark{21,22},
J.~Kamenetzky\altaffilmark{19},
S.~Kim\altaffilmark{13},
M.~Krips\altaffilmark{14},
P.R.~Maloney\altaffilmark{12},
H.~Matsuhara\altaffilmark{23},
A.M.J.~Mortier\altaffilmark{11},
E.J.~Murphy\altaffilmark{6,24},
B.J.~Naylor\altaffilmark{7},
R.~Neri\altaffilmark{14},
H.T.~Nguyen\altaffilmark{6,7},
S.J.~Oliver\altaffilmark{15},
A.~Omont\altaffilmark{18},
M.J.~Page\altaffilmark{25},
A.~Papageorgiou\altaffilmark{2},
C.P.~Pearson\altaffilmark{26,27},
I.~P{\'e}rez-Fournon\altaffilmark{28,29},
M.~Pohlen\altaffilmark{2},
J.I.~Rawlings\altaffilmark{25},
G.~Raymond\altaffilmark{2},
D.~Riechers\altaffilmark{6,30},
G.~Rodighiero\altaffilmark{16},
I.G.~Roseboom\altaffilmark{15},
M.~Rowan-Robinson\altaffilmark{11},
D.~Scott\altaffilmark{31},
N.~Seymour\altaffilmark{25},
A.J.~Smith\altaffilmark{15},
M.~Symeonidis\altaffilmark{25},
K.E.~Tugwell\altaffilmark{25},
M.~Vaccari\altaffilmark{16},
J.D.~Vieira\altaffilmark{6},
L.~Vigroux\altaffilmark{18},
L.~Wang\altaffilmark{15},
G.~Wright\altaffilmark{21},
and J.~Zmuidzinas\altaffilmark{6,7}}
\altaffiltext{1}{Department of Physics and Astronomy, University of Pennsylvania, Philadelphia, PA 19104}
\altaffiltext{2}{School of Physics and Astronomy, Cardiff University, Queens Buildings, The Parade, Cardiff CF24 3AA, UK}
\altaffiltext{3}{Laboratoire AIM-Paris-Saclay, CEA/DSM/Irfu - CNRS - Universit\'e Paris Diderot, CE-Saclay, pt courrier 131, F-91191 Gif-sur-Yvette, France}
\altaffiltext{4}{Department of Physics and Astronomy, Rutgers, The State University of New Jersey, 136 Frelinghuysen Rd, Piscataway, NJ 08854}
\altaffiltext{5}{Institut d'Astrophysique Spatiale (IAS), b\^atiment 121, Universit\'e Paris-Sud 11 and CNRS (UMR 8617), 91405 Orsay, France}
\altaffiltext{6}{California Institute of Technology, 1200 E. California Blvd., Pasadena, CA 91125}
\altaffiltext{7}{Jet Propulsion Laboratory, 4800 Oak Grove Drive, Pasadena, CA 91109}
\altaffiltext{8}{Space Science Building, Cornell University, Ithaca, NY, 14853-6801}
\altaffiltext{9}{Laboratoire d'Astrophysique de Marseille, OAMP, Universit\'e Aix-marseille, CNRS, 38 rue Fr\'ed\'eric Joliot-Curie, 13388 Marseille cedex 13, France}
\altaffiltext{10}{Institute of Astronomy, University of Cambridge, Madingley Road, Cambridge CB3 0HA, UK}
\altaffiltext{11}{Astrophysics Group, Imperial College London, Blackett Laboratory, Prince Consort Road, London SW7 2AZ, UK}
\altaffiltext{12}{Center for Astrophysics and Space Astronomy 389-UCB, University of Colorado, Boulder, CO 80309}
\altaffiltext{13}{Dept. of Physics \& Astronomy, University of California, Irvine, CA 92697}
\altaffiltext{14}{Institut de Radioastronomie Millim\'etrique, 300 Rue de la Piscine, Domaine Universitaire, 38406 Saint Martin d'H\`eres, France}
\altaffiltext{15}{Astronomy Centre, Dept. of Physics \& Astronomy, University of Sussex, Brighton BN1 9QH, UK}
\altaffiltext{16}{Dipartimento di Astronomia, Universit\`{a} di Padova, vicolo Osservatorio, 3, 35122 Padova, Italy}
\altaffiltext{17}{NRAO, PO Box 2, Green Bank, WV 24944}
\altaffiltext{18}{Institut d'Astrophysique de Paris, UMR 7095, CNRS, UPMC Univ. Paris 06, 98bis boulevard Arago, F-75014 Paris, France}
\altaffiltext{19}{Dept. of Astrophysical and Planetary Sciences, CASA 389-UCB, University of Colorado, Boulder, CO 80309}
\altaffiltext{20}{Department of Astronomy, University of Maryland, College Park, MD 20742-2421}
\altaffiltext{21}{UK Astronomy Technology Centre, Royal Observatory, Blackford Hill, Edinburgh EH9 3HJ, UK}
\altaffiltext{22}{Institute for Astronomy, University of Edinburgh, Royal Observatory, Blackford Hill, Edinburgh EH9 3HJ, UK}
\altaffiltext{23}{Institute for Space and Astronautical Science, Japan Aerospace and Exploration Agency, Sagamihara, Kana- gawa 229-8510, Japan}
\altaffiltext{24}{Infrared Processing and Analysis Center, MS 100-22, California Institute of Technology, JPL, Pasadena, CA 91125}
\altaffiltext{25}{Mullard Space Science Laboratory, University College London, Holmbury St. Mary, Dorking, Surrey RH5 6NT, UK}
\altaffiltext{26}{RAL Space, Rutherford Appleton Laboratory, Chilton, Didcot, Oxfordshire OX11 0QX, UK}
\altaffiltext{27}{Institute for Space Imaging Science, University of Lethbridge, Lethbridge, Alberta, T1K 3M4, Canada}
\altaffiltext{28}{Instituto de Astrof{\'\i}sica de Canarias (IAC), E-38200 La Laguna, Tenerife, Spain}
\altaffiltext{29}{Departamento de Astrof{\'\i}sica, Universidad de La Laguna (ULL), E-38205 La Laguna, Tenerife, Spain}
\altaffiltext{30}{Hubble Fellow}
\altaffiltext{31}{Department of Physics \& Astronomy, University of British Columbia, 6224 Agricultural Road, Vancouver, BC V6T~1Z1, Canada}

%% Notice that each of these authors has alternate affiliations, which
%% are identified by the \altaffilmark after each name.  Specify alternate
%% affiliation information with \altaffiltext, with one command per each
%% affiliation.

%\altaffiltext{1}{Visiting Astronomer, Cerro Tololo Inter-American Observatory.
%CTIO is operated by AURA, Inc.\ under contract to the National Science
%Foundation.}
%\altaffiltext{2}{Society of Fellows, Harvard University.}
%\altaffiltext{3}{present address: Center for Astrophysics,
%    60 Garden Street, Cambridge, MA 02138}
%\altaffiltext{4}{Visiting Programmer, Space Telescope Science Institute}
%\altaffiltext{5}{Patron, Alonso's Bar and Grill}

%% Mark off your abstract in the ``abstract'' environment. In the manuscript
%% style, abstract will output a Received/Accepted line after the
%% title and affiliation information. No date will appear since the author
%% does not have this information. The dates will be filled in by the
%% editorial office after submission.

\begin{abstract}
We report on the redshift measurement and CO line excitation of HERMES J105751.1+573027 (HLSW-01), a strongly lensed submillimeter galaxy discovered in {\it Herschel}/SPIRE observations as part of the {\it Herschel} Multi-tiered Extragalactic Survey (HerMES). HLSW-01 is an ultra-luminous galaxy with an {\it intrinsic} far-infrared luminosity of $L_{\mathrm{FIR}} = 1.4 \times 10^{13}$\,$L_\odot$, and is lensed by a massive group of galaxies into at least four images with a total magnification of $\mu = 10.9\pm0.7$. With the 100\,GHz instantaneous bandwidth of the Z-Spec instrument on the Caltech Submillimeter Observatory, we robustly identify a redshift of $z=2.958\pm0.007$ for this source, using the simultaneous detection of four CO emission lines ($J = 7\rightarrow6$, $J = 8\rightarrow7$, $J = 9\rightarrow8$, and $J = 10\rightarrow9$). Combining the measured line fluxes for these high-$J$ transitions with the $J = 1\rightarrow0$, $J = 3\rightarrow2$ and $J = 5\rightarrow4$ line fluxes measured with the Green Bank Telescope, the Combined Array for Research in Millimeter Astronomy, and the Plateau de Bure Interferometer, respectively, we model the physical properties of the molecular gas in this galaxy. We find that the full CO spectral line energy distribution is well described by warm, moderate-density gas with $T_{\mathrm{kin}} = 86-235$\,K and $n_{\mathrm{H}_2} = (1.1-3.5)\times10^3$\,cm$^{-3}$. However, it is possible that the highest-$J$ transitions are tracing a small fraction of very dense gas in molecular cloud cores, and two-component models that include a warm/dense molecular gas phase with $T_{\mathrm{kin}} \sim 200$\,K, $n_{\mathrm{H}_2} \sim 10^5$\,cm$^{-3}$ are also consistent with these data. Higher signal-to-noise measurements of the $J_{\mathrm{up}} \ge 7$ transitions with high spectral resolution, combined with high spatial resolution CO maps, are needed to improve our understanding of the gas excitation, morphology, and dynamics of this interesting high-redshift galaxy.
\end{abstract}

%% Keywords should appear after the \end{abstract} command. The uncommented
%% example has been keyed in ApJ style. See the instructions to authors
%% for the journal to which you are submitting your paper to determine
%% what keyword punctuation is appropriate.


%% From the front matter, we move on to the body of the paper.
%% In the first two sections, notice the use of the natbib \citep
%% and \citet commands to identify citations.  The citations are
%% tied to the reference list via symbolic KEYs. The KEY corresponds
%% to the KEY in the \bibitem in the reference list below. We have
%% chosen the first three characters of the first author's name plus
%% the last two numeral of the year of publication as our KEY for
%% each reference.

%% Authors who wish to have the most important objects in their paper
%% linked in the electronic edition to a data center may do so by tagging
%% their objects with \objectname{} or \object{}.  Each macro takes the
%% object name as its required argument. The optional, square-bracket 
%% argument should be used in cases where the data center identification
%% differs from what is to be printed in the paper.  The text appearing 
%% in curly braces is what will appear in print in the published paper. 
%% If the object name is recognized by the data centers, it will be linked
%% in the electronic edition to the object data available at the data centers  
%%
%% Note that for sources with brackets in their names, e.g. [WEG2004] 14h-090,
%% the brackets must be escaped with backslashes when used in the first
%% square-bracket argument, for instance, \object[\[WEG2004\] 14h-090]{90}).
%%  Otherwise, LaTeX will issue an error. 

\section{Introduction}
\label{sec:intro}

Galaxies selected at submillimeter (submm) and millimeter (mm) wavelengths (hereafter SMGs) are predominantly high-redshift ($1.5 \lesssim z \lesssim 3.5$) systems passing through an important, intense starburst phase of their evolution, with typical star formation rates (SFRs) $>100$\,$M_\odot$\,yr$^{-1}$ \citep[e.g.][]{blain02,chapman05}. Over the past 13 years, a large number of deep, wide-area surveys at $\lambda = 850-1200$\,$\mu$m have been conducted from ground-based telescopes \citep[e.g.][]{scott02,greve04,coppin06,bertoldi07,perera08,weiss09b,austermann10,scott10}. These wavelengths sample the Rayleigh-Jeans tail of thermal dust emission, where this dust is heated by ultraviolet (UV) light from intense star formation or active galactic nuclei (AGNs). Owing to a strong negative $k$-correction, galaxies with the same bolometric luminosity (a proxy for SFR) that are selected at $\lambda \gtrsim 500$\,$\mu$m are equally detectable for $1 \lesssim z \lesssim 10$ in a flux-limited survey. However, the relative ease of detecting a significant number of SMGs in deep, wide-area surveys is countered by the time-consuming multi-wavelength follow-up necessary to derive information on their redshifts, SFRs and star formation efficiencies, morphologies, and dynamics, and on the importance of AGNs in these systems \citep[e.g.][]{chapman05,pope06,younger07,younger09,biggs10}.

Measuring redshifts for a large number of SMGs is necessary for determining their contribution to the cosmic star formation history \citep{chapman05,aretxaga07}, and is crucial for carrying out detailed studies of a representative sample of this population. Obtaining redshifts through optical spectroscopy is difficult, given the poor positional accuracy of most SMGs (arising from the low resolution of ground-based submm/mm telescopes) and their extreme dust obscuration. The most direct redshift measurement for SMGs is through the detection of the rotational transitions of carbon monoxide (CO), since SMGs contain large reservoirs of molecular gas \citep[$\sim10^{10-11}$\,$M_\odot$;][]{greve05,tacconi06,tacconi08}. Until very recently -- with the exception of three blind CO redshift detections \citep{weiss09a,swinbank10,lestrade10} -- CO measurements of SMGs had been limited to sources with known, optically determined redshifts \citep[e.g.][]{frayer98,frayer99,frayer08,ivison01,downes03,genzel03,neri03,sheth04,kneib05,greve05,tacconi06,chapman08,bothwell10,engel10}, since the limited bandwidth and sensitivity of most receivers precluded a blind search for redshifted CO lines.

Two recent developments have spurred the rapid growth of the number of SMGs with blind CO redshift detections over the past year. The first is the availability of sensitive, wide-bandwidth receivers designed specifically for the detection of one or more CO lines. These include Z-Spec, a grating spectrometer with 160 silicon-nitride micro-mesh bolometers operating from $190-310$\,GHz \citep{naylor03,earle06,bradford09}. For $z > 0.5$, at least two CO transitions are redshifted into the Z-Spec bandpass, allowing for a robust redshift determination for galaxies at $z \lesssim 3$ (or higher, provided that the CO $J_{\mathrm{up}} \gtrsim 8$ transitions are excited). The second development is the recent detection of extremely bright SMGs with apparent far-infrared (FIR) luminosities $\gtrsim10^{14}$\,$L_\odot$. Although a few discoveries of such extreme SMGs have been made in small-scale mapping surveys \citep{swinbank10,ikarashi10}, large area ($10-200$\,deg$^2$) surveys carried out with the South Pole Telescope (SPT) at 1.4\,mm \citep{vieira10}, the Atacama Cosmology Telescope at $1-2$\,mm, and with SPIRE \citep[$250-500$\,$\mu$m;][]{griffin10} on the {\it Herschel Space Observatory} \citep{pilbratt10,eales10,oliver10} are uncovering a large number of these extreme SMGs, and are thus providing a sizable number of very bright targets that are ideal for follow-up CO measurements. Five of these extremely bright SMGs that were detected in the {\it Herschel}-ATLAS Science Demonstration Phase data were recently targeted for blind CO measurements with Z-Spec and Zpectrometer on the Green Bank Telescope (GBT), and all five redshifts were successfully measured \citep{lupu10,frayer11}. These redshifts have been key to demonstrating that these extremely bright SMGs are strongly lensed by intervening foreground galaxies \citep{negrello10}. With a growing number of strongly lensed SMGs being uncovered with {\it Herschel}/SPIRE and the SPT, the number of blind CO redshift measurements is expected to grow considerably, even before full operations of the Atacama Large Millimeter/submillimeter Array (ALMA) are underway.

The CO line fluxes constrain the physical properties of molecular gas in a galaxy, including the total molecular gas mass as well as temperature and density. Since estimates of physical conditions depend on the relative line strengths, it is necessary to sample the full spectral line energy distribution (SLED) from the low- to high-$J$ transitions in order to place useful constraints on the gas properties. While most galaxies at both low and high redshifts have only been detected in $1-2$ CO lines \citep{solomon05, greve05, tacconi06, tacconi08, harris10, ivison10a, aravena10, wang10}, a growing number of galaxies have been observed in $\ge3$ transitions in recent years. These include ground-based observations of galaxies at both low and high redshifts \citep{weiss05, riechers06, ao08, weiss07a, weiss07b, bradford09, papadopoulos10a, papadopoulos10b, carilli10, riechers10, lestrade10, danielson10} as well as recent {\it Herschel} SPIRE/Fourier transform spectrometer \citep[FTS;][]{panuzzo10,vanderwerf10} and HIFI \citep{loenen10} observations of nearby starbursts and active galaxies. For many, the full CO SLED is well described by a single component with warm, dense gas typical of star-forming regions, with kinetic temperatures of $T_{\mathrm{kin}} \sim 30-100$\,K and densities of $n_{\mathrm{H}_2} \sim 10^{3.5-6}$\,cm$^{-3}$. However, those with the best sampled SLEDs, spanning the full run of the rotational transition ladder, often require multiple gas phases to explain the observed line fluxes, with a warm/dense phase required to excite the mid- to high-$J$ transitions, and extended, cold low-excitation gas that contributes significantly to the $J_{\mathrm{up}} \lesssim 2$ line fluxes \citep{ward03,carilli10, danielson10, panuzzo10, vanderwerf10}. Several studies have found that most SMGs show an excess in the CO $J=1\rightarrow0$ line luminosity relative to the higher-$J$ transitions \citep{harris10,frayer11}, and high-resolution mapping of CO in a number of these high-redshift galaxies reveals that the $J=1\rightarrow0$ line traces a more extended gas reservoir than that traced by $J_{\mathrm{up}} \gtrsim 3$ transitions \citep{ivison10b}.

In this paper, we present the CO redshift measurement and excitation modeling of HERMES J105751.1+573027 (hereafter HLSW-01), a multiply-imaged SMG discovered in Science Demonstration Phase {\it Herschel}/SPIRE data as part of the {\it Herschel} Multi-Tiered Extragalactic Survey \citep[HerMES\footnote{hermes.sussex.ac.uk};][2011 (in prep.)]{oliver10}. High resolution imaging at $880$\,$\mu$m with the Submillimeter Array (SMA) reveals at least four images separated by $\sim9$\arcsec, and higher resolution $K_p$-band imaging shows strong lensing arcs. The lensing model (Gavazzi et al.~2011; hereafter G11) suggests that HLSW-01 is lensed by a massive group of galaxies, with a total gravitational magnification of $\mu = 10.9\pm0.7$. When corrected for the magnification, this galaxy is found to be a very bright ultra-luminous infrared galaxy (ULIRG) with an FIR luminosity of $L_{\mathrm{FIR}} = (1.43\pm0.09)\times10^{13}$\,$L_\odot$ and an implied SFR of $2460\pm160$\,$M_\odot$\,yr$^{-1}$ (Conley et al.~2011; hereafter C11). The dust continuum spectral energy distribution (SED) is warm with a dust temperature of $T_{\mathrm{d}} \approx 88$\,K determined from a simple one-component modified blackbody model (C11). Combined with a radio flux density in excess of that expected from the FIR to radio correlation, there is tentative evidence that this source harbors a bolometrically important AGN, in which case the SFR inferred from $L_{\mathrm{FIR}}$ is overestimated.

This paper is organized as follows: we discuss the Z-Spec observations and data reduction for HLSW-01 in Section~\ref{sec:obs}; in Section~\ref{sec:zspec}, we describe the redshift determination and line flux measurements from the Z-Spec data; we describe the CO excitation modeling in Section~\ref{sec:radex}, and discuss the molecular gas properties in Section~\ref{sec:disc}. We summarize these results in Section~\ref{sec:conc}. We assume a flat, $\Lambda$CDM cosmology with $\Omega_{\mathrm{M}} = 0.3$, $\Omega_\Lambda = 0.7$, and $H_0 = 70$\,km\,s$^{-1}$\,Mpc$^{-1}$ throughout this paper.

\section{Z-Spec Observations and Data Reduction}
\label{sec:obs}

%% In a manner similar to \objectname authors can provide links to dataset
%% hosted at participating data centers via the \dataset{} command.  The
%% second curly bracket argument is printed in the text while the first
%% parentheses argument serves as the valid data set identifier.  Large
%% lists of data set are best provided in a table (see Table 3 for an example).
%% Valid data set identifiers should be obtained from the data center that
%% is currently hosting the data.
%%
%% Note that AASTeX interprets everything between the curly braces in the 
%% macro as regular text, so any special characters, e.g. "#" or "_," must be 
%% preceded by a backslash. Otherwise, you will get a LaTeX error when you 
%% compile your manuscript.  Special characters do not 
%% need to be escaped in the optional, square-bracket argument.

We carried out the Z-Spec observations of HLSW-01 at the Caltech Submillimeter Observatory (CSO) from 2010 March 09 to May 12 under generally good to excellent observing conditions, with a zenith opacity at 225\,GHz (monitored by the CSO tau meter) ranging over $\tau_{225} = 0.03$ to 0.1, with $\tau_{225} \le 0.06$ for 80\% of the observations. The beamsize ranges from 25\arcsec~to 40\arcsec~(full width at half maximum) over the Z-Spec bandpass. The data were taken using the standard ``chop-and-nod'' mode in order to estimate and subtract the atmospheric signal from the raw data. The secondary mirror was chopped on- and off-source at a rate of 1.6\,Hz, with a chop throw of 90\arcsec\, while stepping through a 4-position nod cycle, integrating for 20\,sec at each nod position. We checked the pointing every $2-4$ hrs by observing quasars and other bright targets located close in elevation to HLSW-01, making small (typically $\lesssim10$\arcsec) adjustments to the telescope pointing model in real time. The total integration time (including the time spent in the off-source position during the nod cycle, but excluding all other overheads) was 22.9 hrs.

We analyze the data using customized software in the same manner as described in \citet{bradford09}. For each channel, the nods are calibrated and averaged together, weighted by the inverse square of the noise. Absolute calibration is determined by frequent ($\sim1$ per night) observations of Mars \citep{wright07} and Neptune\footnote{http://sma1.sma.hawaii.edu/planetvis.html}. The flux density of Mars (proportional to $\nu^2$ from $190-310$\,GHz) decreased with apparent size from $800$\,Jy to $270$\,Jy at 240\,GHz over the course of the Z-Spec Spring 2010 observing season, while the 240\,GHz flux density of Neptune was $16$\,Jy during this run.  We use a total of 42 planet observations taken throughout the observing run to build a model of the flux conversion factor (from voltage to flux density) as a function of operating (``DC'') voltage for each detector separately \citep{bradford09}. Since the DC voltage depends on a combination of the bath temperature and the total optical loading on the detectors, we use these curves to determine appropriate calibration factors to apply to each nod individually. Based on the root mean square (rms) deviations of the planet measurements from the best-fit curves, the channel calibration uncertainties are $4-9$\%, excluding the lowest frequencies for which a good model of the atmosphere is hindered by the 183\,GHz atmospheric water line. These errors are propagated through the data reduction and integrate down as the square root of the number of nods included in the average.

Since blind CO line detections with Z-Spec require small channel-to-channel variations, we determine small bandpass corrections to the calibration for each channel from continuum measurements of bright mm-sources. We carried out a total of 67 observations of bright continuum sources (1\,mm flux density, $S_{1.1} > 1$\,Jy), including J1055+018, J1637+574, J0854+201, 3C273, 3C279, and 3C345. Each was reduced using the calibration determined from the planet observations as described in the previous paragraph. We then fit a power-law to the continuum, separately for each observation, and for each channel compute a multiplicative correction factor to apply to the observed spectrum. We exclude the measurements from the 12 lowest frequency channels ($\nu \le 190$\,GHz, contaminated by the atmospheric water line) and from the 232.3\,GHz channel (known to be unstable) in the continuum fits.  We then average over all observations to determine a single correction factor for each channel. These bandpass corrections are small (ranging from $0.96-1.06$), with little variation ($\le5$\%) over all observations. We apply these small correction factors to the calibrated data to improve the channel-to-channel calibration for our data.

The rms uncertainties on the final co-added spectrum of HLSW-01 range from 1.8 to 4.9\,mJy, or 15\% for most detectors. These errors do not include the uncertainties on the brightness temperatures of Mars and Neptune, which are $\sim5$\%.

\section{Redshift, Continuum, and Line Flux Measurements}
\label{sec:zspec}

\begin{figure*}
\begin{center}
\includegraphics[width=6.5in]{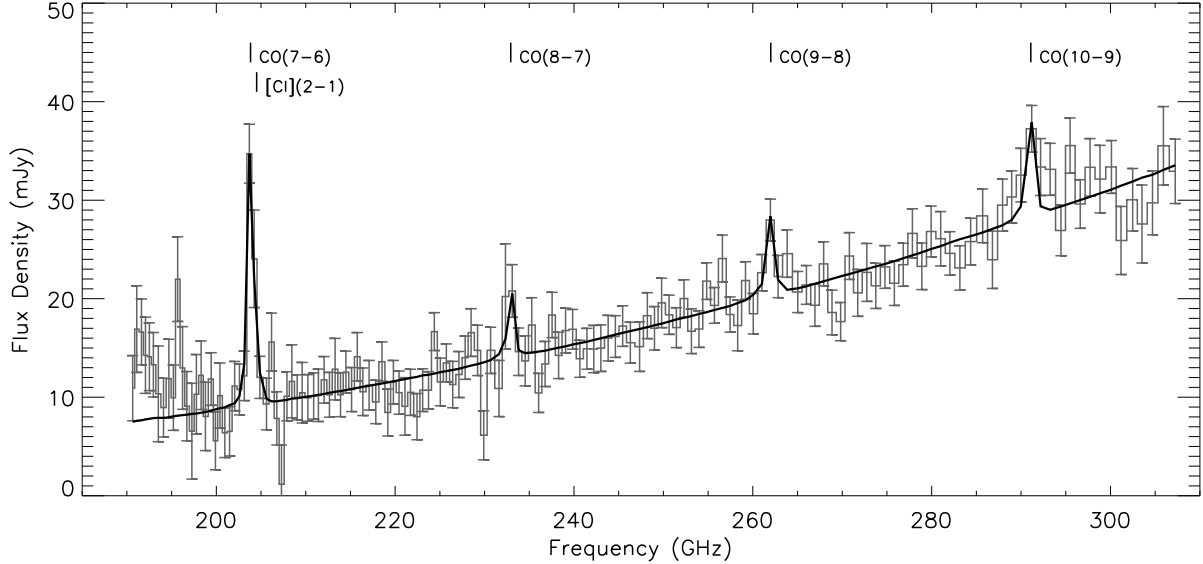}
\caption{The mm-wavelength spectrum for HLSW-01 measured by Z-Spec (histogram). The error bars show the $1\sigma$ photometric errors on the measurements and do not include the 5\% uncertainty on the absolute flux calibration. The solid curve shows the best-fit model to the dust continuum and the CO and [\ion{C}{1}] line emission, with the positions of the lines marked. This model includes the convolution of the intrinsic line profile ($\Delta v = 350$\,km\,s$^{-1}$) with the spectral response profiles of the channels.}
\label{fig:spec} 
\end{center}
\end{figure*}

The $190-310$\,GHz spectrum for HLSW-01 measured by Z-Spec is shown as a histogram in Figure~\ref{fig:spec}. The error bars represent the $1\sigma$ photometric errors on the measurements (not including calibration errors).

\subsection{Redshift Determination}
\label{ssec:zearch}

We use a custom-developed algorithm to determine the redshift of HLSW-01 based on the detection of multiple emission lines. This algorithm is described in detail in \citet[][hereafter L10]{lupu10}, and we summarize it here. Since Z-Spec (with channel widths ranging from 720 to 1290\,km\,s$^{-1}$) does not spectrally resolve the line emission from typical galaxies, the signal from a given line will approximately fall within a single Z-Spec channel, although the spectral response profiles of adjacent channels overlap somewhat \citep{earle06}. Using a reference line list containing all CO transitions up to $J = 17\rightarrow16$ and the fine structure lines from neutral carbon ([\ion{C}{1}]) and singly ionized carbon and nitrogen ([\ion{C}{2}] and [\ion{N}{2}], respectively), we compute two different estimators based on combinations of the continuum-subtracted signal-to-noise ($S/N$) measured in all channels where the reference lines would lie for a given redshift. For HLSW-01, both estimators, $E_1(z)$ and $E_2(z)$ (see L10 for details), are maximal at $z = 2.958\pm0.007$, and with maximum values of $E_{1}(2.958)=9.1$ and $E_{2}(2.958)=5.3$, the redshift is determined with $\gg99.99$\% confidence (L10). Subsequent measurements of the CO $J=5\rightarrow4$ line using the Institut de Radioastronomie Millim\'etrique (IRAM) Plateau de Bure Interferometer (PdBI), the CO $J=3\rightarrow2$ line using the Combined Array for Research in Millimeter Astronomy (CARMA), and the CO $J=1\rightarrow0$ line using the GBT Zpectrometer, confirmed this redshift ($z=2.9574 \pm 0.0001$; Riechers et al. 2011; hereafter R11).

\subsection{Line and Continuum Fitting}
\label{ssec:fit}

We fit the spectrum of HLSW-01 to a model consisting of a power-law continuum and CO line emission. Since the spectrometer outputs only a single value per channel, it is not critically sampled, and so we use the spectral response profile for each channel in the fitting. We exclude in the fit all channels with $\nu \le 190$\,GHz due to poor calibration. We fix all line widths to 350\,km\,s$^{-1}$ (full width at half maximum), which is the line width measured by the PdBI, CARMA, and Zpectrometer assuming a Gaussian profile (R11); however, we note that the fitted line fluxes are fairly insensitive to the choice of line width. The CO $J=7\rightarrow6$ line is separated from the [\ion{C}{1}] $^3${\it P}$_2 \rightarrow ^3${\it P}$_1$ (hereafter [\ion{C}{1}] $2\rightarrow1$) fine structure line by $\sim1000$\,km\,s$^{-1}$, or roughly one Z-Spec channel. For this reason we fix the redshift to $z=2.9574$, the value measured by the PdBI, and include the [\ion{C}{1}] $2\rightarrow1$ line in the fit. The line fluxes (or $5\sigma$ upper limits) for the four CO lines in the Z-Spec bandpass and the [\ion{C}{1}] $2\rightarrow1$ line are shown in Table~\ref{tab:lineflux}, along with the Zpectrometer CO $J=1\rightarrow0$, CARMA CO $J=3\rightarrow2$, and PdBI CO $J=5\rightarrow4$ measurements (R11). We were particularly unlucky in that the $J=8\rightarrow7$ line falls within a noisy channel of the Z-Spec bandpass, so we obtain only an upper limit for this line. The best-fit to the dust continuum is $F_\nu = \left(15.3\pm0.2\right)\,\left({\nu \over 240\,\mathrm{GHz}}\right)^{3.2\pm0.1}$\,mJy. Given a dust temperature of $T_{\mathrm{d}} = 88$\,K determined from fitting the FIR to mm continuum (C11), we are not strictly in the Rayleigh-Jeans limit over the entire Z-Spec bandpass at $z=2.9574$; however, this power-law provides a good fit to these data (reduced $\chi^2$ of $0.97$) and is sufficient for the purposes of baseline fitting. This best-fit model is over-plotted in Figure~\ref{fig:spec}.

\begin{deluxetable}{lrrcccl}
\tabletypesize{\scriptsize}
%\setlength{\tabcolsep}{0.02in}
%\rotate
\tablecaption{CO and [\ion{C}{1}] Transitions Observed in HLSW-01\label{tab:lineflux}}
\tablewidth{0pt}
\tablehead{
\colhead{Transition} & \colhead{$\nu_{\mathrm{rest}}$} & \colhead{$\nu_{\mathrm{obs}}$} & \colhead{Flux Density$^{\mathrm{a}}$}    & \colhead{Line Luminosity$^{\mathrm{a,b}}$}  & \colhead{Line Flux$^{\mathrm{a,b}}$} & \colhead{Reference} \\
\colhead{}           & \colhead{(GHz)}              & \colhead{(GHz)}             & \colhead{(Jy km s$^{-1}$)}              & \colhead{($10^{10}$ K km s$^{-1}$ pc$^2$)} & \colhead{($10^{-20}$ W m$^{-2}$)}        & \colhead{}          \\
}
\startdata
CO~$J=1 \rightarrow 0$ &  115.27 &  29.13 & $ 1.1 \pm 0.1$ & $4.04 \pm 0.39$ & $0.010 \pm 0.001$ & GBT Zpectrometer; R11 \\
CO~$J=3 \rightarrow 2$ &  345.80 &  87.38 & $ 9.7 \pm 0.5$ & $3.83 \pm 0.20$ & $0.26 \pm 0.01$ & CARMA; R11 \\
CO~$J=5 \rightarrow 4$ &  576.27 & 145.61 & $23.6 \pm 1.4$ & $3.34 \pm 0.20$ & $1.05 \pm 0.06$ & IRAM PdBI; R11 \\
CO~$J=7 \rightarrow 6$ &  806.65 & 203.83 & $34.6 \pm 4.2$ ($^{+ 9.9}_{-8.1}$) & $2.50 \pm 0.30$ ($^{+0.72}_{-0.58}$) & $2.2 \pm 0.3$ ($^{+0.6}_{-0.5}$) & Z-Spec; this work \\
CO~$J=8 \rightarrow 7$ &  921.80 & 232.92 & $< 20$ ($< 30$) & $< 1.1$ ($< 1.7$) & $< 1.5$ ($< 2.2$) & Z-Spec; this work \\
CO~$J=9 \rightarrow 8$ & 1036.91 & 262.01 & $12.7 \pm 3.3$ ($^{+ 3.7}_{-3.3}$) & $0.56 \pm 0.15$ ($^{+0.16}_{-0.14}$) & $1.0 \pm 0.3$ ($^{+0.3}_{-0.3}$) & Z-Spec; this work \\
CO~$J=10 \rightarrow 9$ & 1151.99 & 291.09 & $14.7 \pm 3.7$ ($^{+ 4.9}_{-4.1}$) & $0.52 \pm 0.13$ ($^{+0.17}_{-0.14}$) & $1.3 \pm 0.3$ ($^{+0.4}_{-0.4}$) & Z-Spec; this work \\
$[$\ion{C}{1}$]$~$^3${\it P}$_2 \rightarrow ^3${\it P}$_1$ &  809.34 & 204.51 & $< 25$ ($< 48$) & $< 1.8$ ($< 3.5$) & $< 1.6$ ($< 3.0$) & Z-Spec; this work \\
\enddata
\tablenotetext{a}{For the Z-Spec data, the initial errors and $5\sigma$ upper limits come from the formal statistical errors from fitting the spectrum. The errors and $5\sigma$ upper limits in parentheses include the uncertainty in the Z-Spec frequency scale as described in Section~\ref{ssec:fit}.}
\tablenotetext{b}{Corrected for magnification assuming $\mu = 10.9$.}
\end{deluxetable}

The formal errors on the line fluxes derived from the above fit do not include the uncertainty of the Z-Spec frequency scale, which is $\sigma_v\sim100$\,km\,s$^{-1}$. This has important implications for the CO $J=7\rightarrow6$ and [\ion{C}{1}] $2\rightarrow1$ line fluxes, as these lines are blended. We include this additional uncertainty on the lines fluxes by shifting the redshift by $\pm\sigma_z = \pm(1+z)\,{\sigma_v/c} = \pm0.001$, refitting the line fluxes, and taking the upper and lower bounds from the statistical $1\sigma$ errors for these fits. These $\pm1\sigma$ uncertainties in the best-fit fluxes are listed in parentheses in Table~\ref{tab:lineflux}. For the CO $J=7\rightarrow6$ and [\ion{C}{1}] $2\rightarrow1$ lines, including the frequency scale uncertainty roughly doubles the error bounds on the measured lines fluxes.

The CO $J = 10\rightarrow9$ line appears somewhat broader than the other transitions and is brighter than the $J = 9\rightarrow8$ line; we therefore consider possible blending with the o-H$_2$O $3_{12}\rightarrow2_{21}$ water line at $\nu_{\mathrm{rest}} = 1153.13$\,GHz. Recent observations of submm-bright AGN-host galaxies show evidence for water vapor emission \citep{bradford09,lupu10}, the most striking example being the local ULIRG Mrk 231, where the {\it Herschel} SPIRE/FTS spectrum reveals seven rotational emission lines of water \citep{gonzalezalfonso10}. Given the close proximity in frequency of the CO $J = 10\rightarrow9$ and o-H$_2$O $3_{12}\rightarrow2_{21}$ lines, we cannot deblend these lines within the Z-Spec spectrum; however, theoretical arguments supported by observed water line ratios in nearby galaxies suggest that the CO $J = 10\rightarrow9$ line flux for HLSW-01 is not contaminated by water emission. The o-H$_2$O $3_{12}\rightarrow2_{21}$ line is part of a de-excitation cascade process, following the excitation of the $3_{21}$ level through the o-H$_2$O $2_{12}\rightarrow3_{21}$ 75~$\mu$m transition. The subsequent cascade results in emission of the o-H$_2$O $3_{21}\rightarrow3_{12}$ ($\nu_{\mathrm{rest}} = 1162.91$\,GHz), o-H$_2$O $3_{12}\rightarrow3_{03}$ ($\nu_{\mathrm{rest}} = 1097.37$\,GHz), and o-H$_2$O $3_{12}\rightarrow2_{21}$ lines \citep[e.g., see Figure 2 in][]{gonzalezalfonso10}. Photon number conservation implies that the sum of the photons emitted in the o-H$_2$O $3_{12}\rightarrow2_{21}$ and o-H$_2$O $3_{12}\rightarrow3_{03}$ lines equals the number of photons emitted in the o-H$_2$O $3_{21}\rightarrow3_{12}$ line. Since neither the o-H$_2$O $3_{21}\rightarrow3_{12}$ nor the o-H$_2$O $3_{12}\rightarrow3_{03}$ lines are detected in the spectrum of HLSW-01 at the $1\sigma$ level, we argue that the strength of the o-H$_2$O $3_{12}\rightarrow2_{21}$ line is also consistent with the noise in our measurements. Furthermore, the observed o-H$_2$O $3_{21}\rightarrow3_{12}$ and p-H$_2$O $2_{02}\rightarrow1_{11}$ ($\nu_{\mathrm{rest}} = 987.92$\,GHz) transitions are stronger than the o-H$_2$O $3_{12}\rightarrow2_{21}$ line for both Mrk 231 \citep{gonzalezalfonso10} and Arp 220 (N. Rangwala, private communication). This adds observational evidence against significant blending of the CO $J=10\rightarrow9$ line with the o-H$_2$O $3_{12}\rightarrow2_{21}$ line in HLSW-01, since we do not detect the stronger water lines in the spectrum. 

\begin{figure}
\begin{center}
\includegraphics[width=3in]{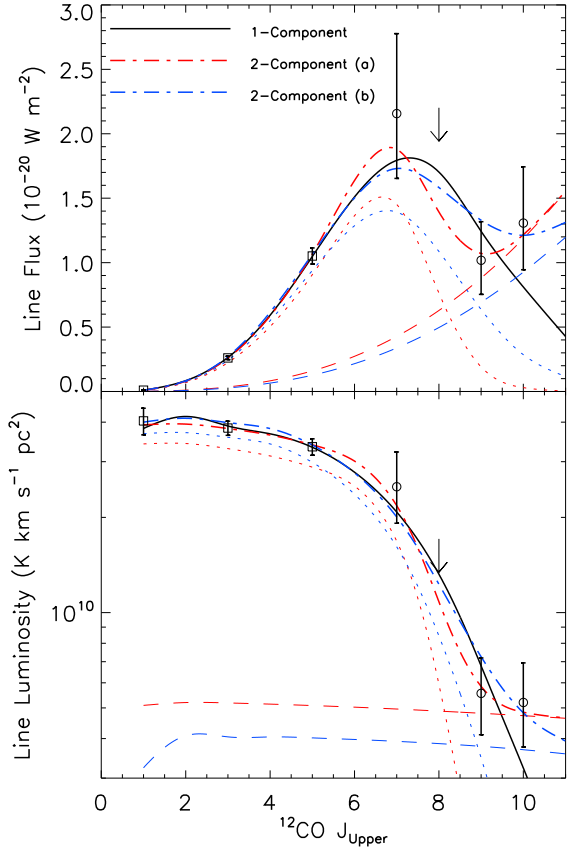}
\caption{The CO spectral line energy distribution for HLSW-01, in line flux (top) and line luminosity units (bottom). The circles and upper limit ($5\sigma$) are from the Z-Spec measurements (this work), where the $1\sigma$ error bars include the uncertainty from the Z-Spec frequency scale. The squares are from the GBT (CO $J=1\rightarrow0$), CARMA (CO $J=3\rightarrow2$), and PdBI (CO $J=5\rightarrow4$) observations (R11). All data have been corrected for magnification assuming $\mu = 10.9$. The black solid curve represents the maximum likelihood single-component model discussed in Section~\ref{ssec:onecomp}. The red dot-dashed curve shows the maximum likelihood solution from a two-component gas model described in Section~\ref{ssec:twocomp}. The red dashed and dotted curves show the contribution from the warm and cold components, respectively. The blue dot-dashed curve demonstrates a different two-component gas model (also described in Section~\ref{ssec:twocomp}) that provides a good fit to these data; in this case, both components (blue dashed and dotted curves) are relatively warm.} 
\label{fig:sled} 
\end{center}
\end{figure}

The CO SLED in units of W\,m$^{-2}$ and in line luminosity units \citep[$L^\prime_{\mathrm{CO}}$,][]{solomon05} is shown in Figure~\ref{fig:sled}. These values have been corrected for a magnification of $\mu = 10.9$, as determined from the lensing model of G11. The SLED turns over at $6 \lesssim J \lesssim 8$, as is typical for SMGs and quasars \citep{weiss07a, weiss07b, danielson10}; however, it is not possible to identify the turnover precisely given that the CO $J=7\rightarrow6$ line flux may be overestimated due to blending with the [\ion{C}{1}] $2\rightarrow1$ emission line, and we have only an upper limit on the CO $J=8\rightarrow7$ line flux. 

\section{Excitation and Radiative Transfer Modeling}
\label{sec:radex}

The SLED for HLSW-01 is well sampled from low- to high-$J$, and thus allows for a rigorous analysis of the CO line excitation. We use RADEX, a non-LTE radiative transfer code which uses an iterative escape probability formalism \citep{vandertak07}, to model the CO excitation of HLSW-01. For a specified gas density ($n_{\mathrm{H}_2}$), kinetic temperature ($T_{\mathrm{kin}}$), and CO column density per unit line width ($N_{\mathrm{CO}}/\Delta v$), RADEX calculates the excitation temperatures, line optical depths, and line surface brightnesses. For our analysis, we use the escape probability in an expanding spherical cloud and assume a $T_{\mathrm{CMB}}(z) = 2.73\,(1+z)$\,K blackbody for the background radiation field. However, we note that the RADEX results are not sensitive to the precise form of escape probability chosen.

Following a similar analysis to that described in \citet{ward03} \citep[see also][and Kamenetzky et al. 2011]{bradford09,naylor10}, we use RADEX to compute the expected CO line intensities over parameter space in $T_{\mathrm{kin}}$, $n_{\mathrm{H}_2}$, and $N_{\mathrm{CO}}$, assuming a line width of $\Delta v = 350$\,km\,s$^{-1}$ (R11) for all transitions. Rather than computing line intensities over a grid in parameter space, we use a Markov-chain Monte Carlo method \citep{metropolis53,hastings70} to determine the likelihood distributions for these parameters by comparing the RADEX results to the observed line intensities. The integrated line fluxes, $S_{\mathrm{CO}}\Delta v$ (in Jy\,km\,s$^{-1}$, Table~\ref{tab:lineflux}) are converted into Rayleigh-Jeans equivalent velocity-integrated intensities, $T_{\mathrm{R}}\Delta v$ (in K\,km\,s$^{-1}$) for comparison with the output from RADEX:

\begin{equation}
\label{eqn:flux2RJ}
T_{\mathrm{R}}\Delta v = {c^2 \over 2\,k\,\nu_{\mathrm{obs}}^2} {S_{\mathrm{CO}}\Delta v \over \mu\,\Omega_{\mathrm{em}}} (1+z),
\end{equation}

\noindent where $\mu = 10.9$ (G11) is the gravitational lensing magnification factor, and $\Omega_{\mathrm{em}}$ is the {\it intrinsic} solid angle of the CO emitting region (see the next paragraph). We propagate the line flux errors using the $1\sigma$ error bounds that include the uncertainty in the Z-Spec frequency scale. We do not include errors in the absolute flux calibration as these are small compared to the random errors on the measurements; nor do we account for systematic calibration errors among the measurements, as these are not possible to characterize without at least two measurements of the same CO transition using different telescopes. 

The CO emission is only marginally resolved in both the IRAM PdBI and CARMA maps. Given the uncertainties in the size of the CO emitting region and the gravitational magnification, combined with the fact that the area filling factor of the molecular gas may be $<1$, we choose to include the intrinsic source solid angle, $\Omega_{\mathrm{em}}$, as a fourth parameter in our likelihood analysis. The intrinsic solid angle refers to the observed solid angle in the absence of lensing. Finally, we impose an upper limit for the kinetic temperature of 3000\,K; above this temperature, collisional dissociation of CO becomes important (with a weak dependence on the gas density).

\begin{figure*}
\begin{center}
\includegraphics[width=6.5in]{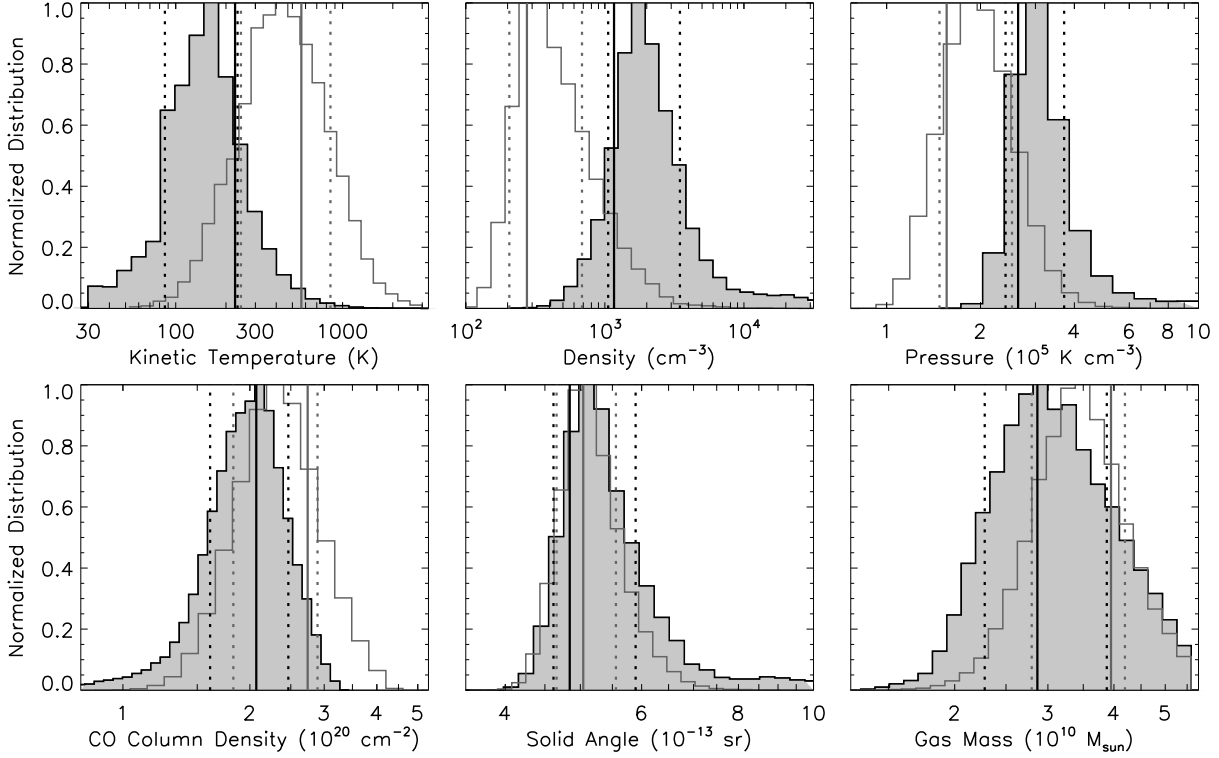}
\caption{{\it Top left, top center, bottom left, bottom center:} Marginalized distributions for the four parameters from the excitation modeling of HLSW-01. {\it Top right, bottom right:} Marginalized distributions for pressure and gas mass, derived from the four primary parameters. All of the distributions are normalized at their peaks. In each panel, the gray, unshaded histogram shows the resulting distribution when the gas is not required to be self-gravitating, while the shaded histogram shows the distribution when this constraint is enforced (i.e. Equation~\ref{eqn:kvir}). The vertical solid and dotted lines indicate the maximum likelihood solutions and the marginalized 68.3\% confidence intervals, respectively, with (black) and without (gray) the constraint of self-gravitating gas. The source solid angle and gas mass distributions assume a gravitational lensing magnification factor of $\mu = 10.9$ for HLSW-01. For a different value of $\mu$, multiply the {\it x}-axes in the bottom center and right plots by $\left({10.9 \over \mu}\right)$.}
\label{fig:1dprim} 
\end{center}
\end{figure*}

\begin{figure*}
\begin{center}
\includegraphics[width=6.5in]{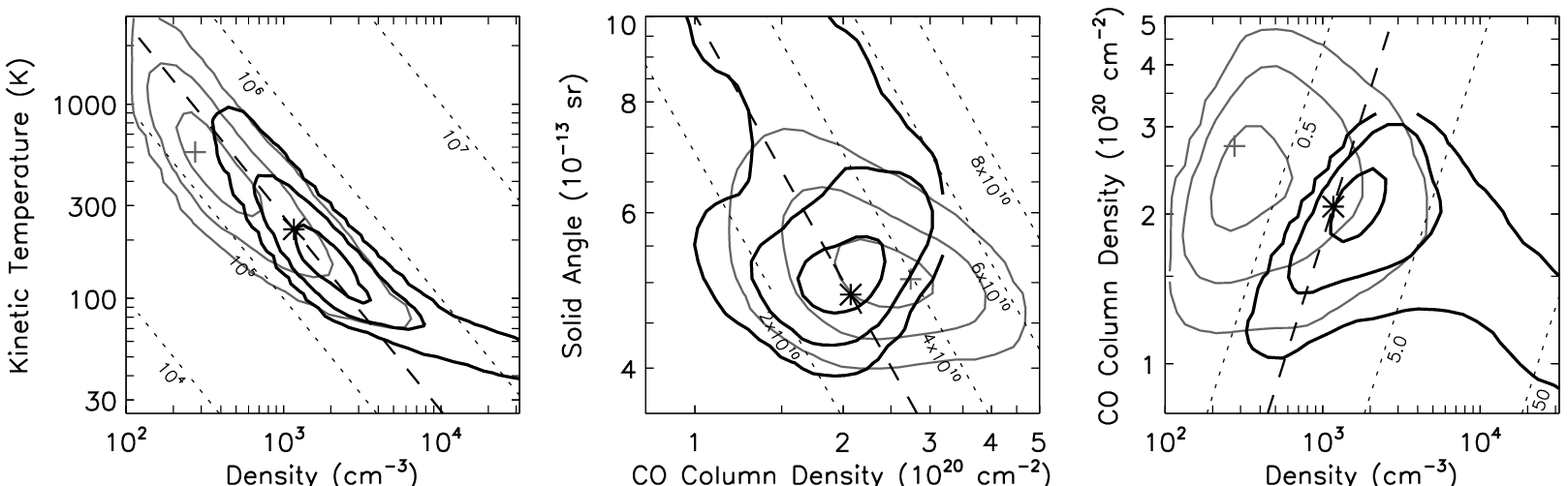}
\caption{The 2-dimensional marginalized distributions for $T_{\mathrm{kin}}$ and $n_{\mathrm{H}_2}$ (left), $\Omega_{\mathrm{em}}$ and $N_{\mathrm{CO}}$ (center), and $N_{\mathrm{CO}}$ and $n_{\mathrm{H}_2}$ (right). In all plots, the contours are 68.3\%, 95.5\%, and 99.7\% assuming Gaussian distributed errors. The gray contours show the resulting distributions when the gas is not constrained to be self-gravitating, and the gray plus symbols mark the maximum likelihood solution in that case. The black contours show the distributions with the constraint that the gas is self-gravitating, with the black stars marking the maximum likelihood solution. In the left plot, the dotted lines indicate lines of constant pressure (in units of K\,cm$^{-3}$), and the dashed pressure line passes through the maximum likelihood solution. In the center plot, the dotted lines show constant gas mass (in units of $M_\odot$), and the dashed line passes through the maximum likelihood solution. The dotted lines in the right plot show constant velocity gradient (in units of km\,s$^{-1}$\,pc$^{-1}$), where the dashed line indicates the maximum likelihood solution. The source solid angle assumes a gravitational lensing magnification of $\mu = 10.9$ for HLSW-01. For a different value of $\mu$, multiply the {\it y}-axes in the center plot by $\left({10.9 \over \mu}\right)$.}
\label{fig:2dprim} 
\end{center}
\end{figure*}

We ran a $150,000$-step Markov chain to determine the likelihood distributions of the gas parameters for HLSW-01. The marginalized distributions for the four parameters are shown in Figure~\ref{fig:1dprim} (gray, unshaded histogram), with the maximum likelihood values marked with gray solid vertical lines. Note that since these are marginalized distributions, the peaks do not necessarily coincide with the 4-dimensional maximum likelihood solution. We use these 1-dimensional distributions only for computing the marginalized 68.3\% confidence interval on each of the parameters, which is indicated by the gray vertical dotted lines in Figure~\ref{fig:1dprim}. These results are summarized in Table~\ref{tab:maxlike}. Figure~\ref{fig:2dprim} shows the 2-dimensional marginalized distributions for $T_{\mathrm{kin}}$ and $n_{\mathrm{H}_2}$, $\Omega_{\mathrm{em}}$ and $N_{\mathrm{CO}}$, and $N_{\mathrm{CO}}$ and $n_{\mathrm{H}_2}$ (gray contours). For $T_{\mathrm{kin}}$ and $n_{\mathrm{H}_2}$, the contours fall along a line of constant pressure.

We also derive the gas pressure ($P = n_{\mathrm{H}_2}\,T_{\mathrm{kin}}$) and mass, where the latter is estimated as:

\begin{equation}
\label{eqn:mgas}
M_{\mathrm{gas}} = {\Omega_{\mathrm{em}}\,D_{\mathrm{A}}^2 N_{\mathrm{CO}} (1.4\,m_{\mathrm{H}_2}) \over X_{\mathrm{CO}}},
\end{equation}

\noindent where $\Omega_{\mathrm{em}}\,D_{\mathrm{A}}^2$ is the area of the emitting region, $D_{\mathrm{A}}$ is the angular diameter distance, $X_{\mathrm{CO}} = n_{\mathrm{CO}} / n_{\mathrm{H}_2}$ is the relative abundance of CO to H$_2$, and $m_{\mathrm{H}_2}$ is the mass of the hydrogen molecule (with a factor of 1.4 to account for helium). We assume $X_{\mathrm{CO}} = 2\times10^{-4}$, a reasonable value based on observations of nearby galaxies and star-forming regions in the Galaxy \citep[e.g.][]{wang04}. The marginalized distributions for these secondary parameters are also shown in Figure~\ref{fig:1dprim}.

\begin{deluxetable}{lccc}
\tabletypesize{\scriptsize}
%\setlength{\tabcolsep}{0.02in}
%\rotate
\tablecaption{Results of Single-phase CO Excitation Modeling for HLSW-01\label{tab:maxlike}}
\tablewidth{0pt}
\tablehead{
\colhead{                   } & \colhead{$K_{\mathrm{vir}}$ unconstrained} & \colhead{$K_{\mathrm{vir}} \ge 1$}                         & \colhead{         } \\
\colhead{Parameter} & \colhead{Maximum Likelihood Solution}         & \colhead{Maximum Likelihood Solution} & \colhead{Units} \\
}
\startdata
$T_{\mathrm{kin}}$ & $566$ ($246 - 845$ ) & $227$ ($ 86 - 235$ ) & K \\
$n_{\mathrm{H}_2}$ & $0.3$ ($0.2 -  0.7$ ) & $1.2$ ($1.1 -  3.5$ ) & $10^3$ cm$^{-3}$ \\
$N_{\mathrm{CO}}$ & $2.7$ ($1.8 - 2.9$ ) & $2.1$ ($1.6 - 2.5$ ) & $10^{20}$ cm$^{-2}$ \\
$\Omega_{\mathrm{em}}$ & $5.0$ ($4.7 - 5.6$ ) & $4.8$ ($4.6 - 5.9$ ) & $\left({10.9 \over \mu}\right)10^{-13}$ sr \\
\hline
$P$ & $1.6$ ($1.5 - 2.5$ ) & $2.6$ ($2.4 - 3.7$ ) & $10^5$ K cm$^{-3}$ \\
$M_{\mathrm{gas}}$ & $3.9$ ($2.8 - 4.2$ ) & $2.9$ ($2.3 - 3.9$ ) & $\left({2\times10^{-4} \over X_{\mathrm{CO}}}\right)\left({10.9 \over \mu}\right)10^{10}$ M$_\odot$ \\
$dv/dr$ & $0.22$ ($0.18 - 0.55$ ) & $1.2$ ($1.1 - 3.0$ ) & $\left({X_{\mathrm{CO}} \over 2\times10^{-4}}\right)$ km s$^{-1}$ pc$^{-1}$ \\
\enddata
\tablecomments{The columns are as follows: 1) the parameter; 2) the maximum likelihood value of the parameter and 68.3\% confidence interval (in parenthesis), without the constraint that the gas is self-gravitating; 3) the maximum likelihood value of the parameter and 68.3\% confidence interval when the gas is required to be self gravitating (Equation~\ref{eqn:kvir}); and 4) the units for the given values.}
\end{deluxetable}

The maximum likelihood solution gives $T_{\mathrm{kin}} \approx 570$\,K and $n_{\mathrm{H}_2} \approx 3\times10^2$\,cm$^{-3}$, a condition that is unlike the average molecular gas seen in most galaxies. This solution is also inconsistent with a gas that has at least enough velocity dispersion to correspond to virialized motion under its own self gravity. Using the notation of \citet{papadopoulos07} and reforming their Equation~2 for our parameterization yields the following constraint for a self-gravitating gas:

\begin{equation}
\label{eqn:kvir}
K_{\mathrm{vir}} \equiv {(dv/dr)_{\mathrm{obs}} \over (dv/dr)_{\mathrm{vir}}} = {\Delta v\,n_{\mathrm{H}_2}^{1/2}\,X_{\mathrm{CO}} \over (4 \pi G \alpha (1.4\,m_{\mathrm{H}_2}) / 3)^{1/2}\,N_{\mathrm{CO}}} \ge 1
\end{equation}

\noindent where $\alpha = 1-2.5$ depending on the cloud density profile, and we estimate the velocity gradient for an individual cloud as $(dv/dr)_{\mathrm{obs}} = \left({\Delta v \over N_{\mathrm{CO}}}\right)\,n_{\mathrm{H}_2}\,X_{\mathrm{CO}}$ (Kamenetzky et al. 2011). Assuming $X_{\mathrm{CO}} = 2\times10^{-4}$, $K_{\mathrm{vir}} = 0.2-0.4$ for the maximum likelihood solution. For this reason, we add a prior to our likelihood analysis that constrains parameter space to solutions where the gas is self-gravitating, i.e. where Equation~\ref{eqn:kvir} holds, and we repeat our Markov chain calculations. These results are presented in Table~\ref{tab:maxlike}, Figure~\ref{fig:1dprim} (shaded histograms), and Figure \ref{fig:2dprim} (black contours), where the maximum likelihood solution now gives $T_{\mathrm{kin}} \approx 230$\,K and $n_{\mathrm{H}_2} \approx 1.2\times10^3$\,cm$^{-3}$. Throughout the rest of this paper, we discuss only the solution where the constraint on $K_{\mathrm{vir}}$ has been enforced.

\section{Results and Discussion}
\label{sec:disc}

Our likelihood analysis for HLSW-01 suggests that the CO gas is primarily tracing warm, moderate-density gas with $T_{\mathrm{kin}} = 227^{+8}_{-141}$\,K and $n_{\mathrm{H}_2} = 1.2^{+2.3}_{-0.1}\times10^3$\,cm$^{-3}$. This single-component model provides a good fit to our data, agreeing within $1.5\sigma$ of all measurements as demonstrated in Figure~\ref{fig:sled}, where the model line fluxes corresponding to our maximum likelihood solution (solid black curve) are compared to our data. However, this model falls short of the $J = 10\rightarrow9$ line flux, and it is likely that the full CO SLED is really a composite of multiple gas components, as seen in many nearby starbursts and high-redshift galaxies \citep{ward03,weiss07a, greve09, panuzzo10, loenen10, riechers10, danielson10, carilli10, vanderwerf10}. In Section~\ref{ssec:onecomp}, we discuss the implications our single-component gas model would have for HLSW-01, and in Section~\ref{ssec:twocomp}, we consider a few two-phase gas models that are also consistent with our data.

\subsection{Single-component Gas Model}
\label{ssec:onecomp}

For our best-fit single-component gas model, all of the CO transitions are sub-thermally populated, with an excitation temperature of $T_{\mathrm{ex}} \sim 135$\,K for the CO $J=1\rightarrow0$ line, and decreasing to $T_{\mathrm{ex}} \sim 25$\,K for the CO $J=10\rightarrow9$ line. The high gas kinetic temperature ($T_{\mathrm{kin}} = 230$\,K) is required to excite the higher-$J$ transitions, and is considerably larger than the dust temperature of $T_{\mathrm{d}} = 88$\,K determined from C11. This suggests that heating mechanisms that effectively transfer energy into the gas (e.g. X-ray dominated regions, or heating by the dissipation of turbulence or cosmic rays) may be present in this system (although colder gas solutions that match the dust temperature are also consistent with our data). 

Our result for HLSW-01 is similar to that for the local starburst galaxy M82, where the $J_{\mathrm{up}} \ge 4$ lines are best-fit by warm gas with moderate density \citep[$T_{\mathrm{kin}} \approx 550$\,K, $n_{\mathrm{H}_2} \approx 5\times10^3$\,cm$^{-3}$,][]{panuzzo10}. However, in contrast to M82, this warm gas is able to account for the excitation of the lower-$J$ lines as well, whereas in M82 the $J_{\mathrm{up}} \le 3$ line fluxes are dominated by a cold, less dense component \citep{wild92,guesten93,mao00,ward03,weiss05b}. Given its low critical density, the CO $J=1\rightarrow0$ line is easily excited in both cold, diffuse gas as well as warm, dense gas in star-forming regions. There is growing evidence that many high-redshift SMGs contain a significant amount ($\sim$50\%  of the total mass) of extended, cold gas that is not associated with the nuclear starburst \citep[e.g.][]{ivison10b,danielson10,carilli10}. This leads to an excess in the CO $J=1\rightarrow0$ line luminosity relative to the higher-$J$ transitions, with a line ratio of $R_{3,1} \equiv L_{\mathrm{CO(3-2)}}^\prime/L_{\mathrm{CO(1-0)}}^\prime \sim 0.6$ \citep[e.g.][]{harris10,frayer11,ivison10b} and a line width ratio of $\Delta v_{\mathrm{CO(1-0)}}/\Delta v_{\mathrm{CO(3-2)}} = 1.15\pm0.06$ \citep{ivison10b}. However, this does not appear to be the case for HLSW-01, where $R_{3,1} = 0.95$ and the observed line widths for the $J_{\mathrm{up}} = 1$, 3, and 5 transitions are all consistent with the presence of a single component. This suggests that the $J_{\mathrm{up}} \le 5$ transitions for HLSW-01 are largely tracing the same volume of molecular gas in this galaxy, and given the large uncertainties in the higher-$J$ observations, we see no strong evidence that a two-phase gas model is needed to explain the CO excitation for this source. 

We derive a source solid angle of $\Omega_{\mathrm{em}} = 4.8^{+1.1}_{-0.2}\left({10.9 \over \mu}\right)\times10^{-13}$\,sr, which corresponds to an equivalent diameter of $1.25^{+0.14}_{-0.03}\left({10.9 \over \mu}\right)^{1/2}$ kpc at $z = 2.9574$, assuming spherical geometry. Combined with the warm temperature and modest density, this suggests that the molecular gas is excited by an intense starburst that is less centrally concentrated than that seen in local ULIRGs, but consistent with previous findings of extended star formation in SMGs \citep[$1.5-3$\,kpc, e.g.][]{biggs08,tacconi08,lestrade10,swinbank10}. It is possible that the molecular gas reservoir is extended over an even larger area \citep[e.g.,][]{ivison10b,carilli10}, and $\Omega_{\mathrm{em}}$ really represents the product of this area with a filling fraction. The velocity gradients observed for the four images in the CO $J=3\rightarrow2$ and CO $J=5\rightarrow4$ maps suggest a complex velocity structure, possibly arising from a major merger (R11); however, higher spatial resolution CO maps are needed in order to confirm this interpretation.

The total molecular gas mass derived from this line excitation modeling (Equation~\ref{eqn:mgas}) is $2.9^{+1.0}_{-0.6}\left({2\times10^{-4} \over X_{\mathrm{CO}}}\right)\left({10.9 \over \mu}\right) \times 10^{10}$\,$M_\odot$, similar to those of other high-redshift SMGs and quasars \citep[e.g.][]{tacconi08,coppin08,wang10,riechers10}. This value is consistent with the estimate from $L^\prime_{\mathrm{CO}(1-0)}$ ($3.3\times10^{10}$\,$M_\odot$, R11) assuming a conversion factor of $\alpha_{\mathrm{CO}} = 0.8$\,$M_\odot$\,(K km s$^{-1}$ pc$^2$)$^{-1}$, a value appropriate for ultra-luminous infrared galaxies \citep{downes98,solomon05}. C11 derive an SFR of $2460\pm160$\,$M_\odot$\,yr$^{-1}$ (corrected for lensing) from the FIR luminosity \citep{kennicutt98}. Assuming that all of the molecular gas will be consumed by star formation, this implies a gas depletion timescale of $11.7^{+4.2}_{-2.5}$\,Myr --- a value similar to \citep[12\,Myr,][]{ivison10b} or somewhat shorter than \citep[$\sim40$\,Myr][]{greve05,solomon05} the mean gas depletion timescale of typical SMGs with similar luminosities. However, it is possible that the SFR is overestimated if a significant fraction of the dust is heated by AGN activity, and this estimate also neglects feedback mechanisms.

At the resolution of the existing interferometry measurements, the SMA 880\,$\mu$m and CO $J=3\rightarrow2$ and $J=5\rightarrow4$ maps show consistent source sizes, suggesting that the dust and gas are well mixed and therefore trace the same regions. We estimate the dust mass from the optical depth and the dust absorption coefficient, where

\begin{equation}
\label{eqn:dustmass}
M_{\mathrm{dust}} = \Omega_{\mathrm{em}}\,D_{\mathrm{A}}^2 {\tau_\nu \over \kappa_\nu},
\end{equation}

\noindent $\kappa_\nu = 2.64\,\mathrm{m}^2\,\mathrm{kg}^{-1} (\nu/2400\mathrm{GHz})^\beta$ \citep{dunne03}, and $\tau_\nu = (\nu/\nu_0)^\beta$, with $\beta = 1.94 \pm 0.14$ and $\nu_0 = 1550 \pm 150$\,GHz determined from fitting a modified blackbody to the full FIR-mm dust SED of HLSW-01 (C11). We derive a dust mass of $M_{\mathrm{d}} = 5.2^{+1.6}_{-1.1} \times 10^8$\,$M_\odot$ (not including errors on $\kappa_\nu$), and a dust to gas mass ratio of $\lesssim1/55$, considering that the total gas mass could be larger since we only estimate the mass of the molecular gas. This is consistent with the dust to gas mass ratios of typical SMGs \citep[$\sim1/60$; e.g.][]{kovacs06,coppin08,michalowski10,santini10}. Our dust mass estimate is $\sim5$ times higher than that estimated via $L_{\mathrm{FIR}}$ from the single-component dust model of C11, assuming the same $\kappa_\nu$. This suggests that 1) a single-component model for the dust (and/or CO) emission is not a good representation of the physical properties of this galaxy, and/or 2) the gas and dust are not co-spatial. Longer-wavelength continuum measurements are needed to detect the presence of multi-component dust, while higher spatial resolution imaging of the dust and CO would allow us to determine their relative spatial distribution. 

Recent results by \citet{papadopoulos10b} suggest that for many (U)LIRGs, including Arp 220 \citep{papadopoulos10a}, high dust extinction can suppress the high-$J$ CO line fluxes. Assuming a single-component dust model for HLSW-01 (C11), the dust optical depths at the CO $J_{\mathrm{up}} = 7-10$ rest frequencies are $\tau_\nu \sim 0.3-0.6$, and thus may be non-negligible. However, we cannot definitively know whether this effect is important for HLSW-01 given the degeneracy between high dust optical depths and large amounts of cold dust. A two-temperature component fit to the dust SED is poorly constrained and does not represent an improvement over a single-component model (C11). However, this fit only includes rest-frame frequencies $\nu \gtrsim 900$\,GHz, and it is possible that a significant amount of cold dust which is best traced at lower frequencies \citep{dunne00,dunne01} is present in this galaxy. Furthermore, to properly include this effect in the excitation modeling would require knowledge of the relative geometry of the molecular gas and dust, which cannot be deduced from our current data. High resolution mapping of the molecular gas and dust emission may shed light on this issue in the future.

\subsection{Two-component Gas Models}
\label{ssec:twocomp}

The single-component gas model discussed in the previous section provides a good fit to the CO SLED for HLSW-01. However, in light of recent results showing that the CO excitation for a large number of local starbursts and high-redshift SMGs requires multiple gas phases, we explore a few two-component gas models in this section. With only six line fluxes measured, we cannot well constrain a full 8-parameter fit; however, we can place simple constraints on some of the parameters in order to test solutions that resemble those of other high-redshift SMGs.

\begin{figure}
\begin{center}
\includegraphics[width=3in]{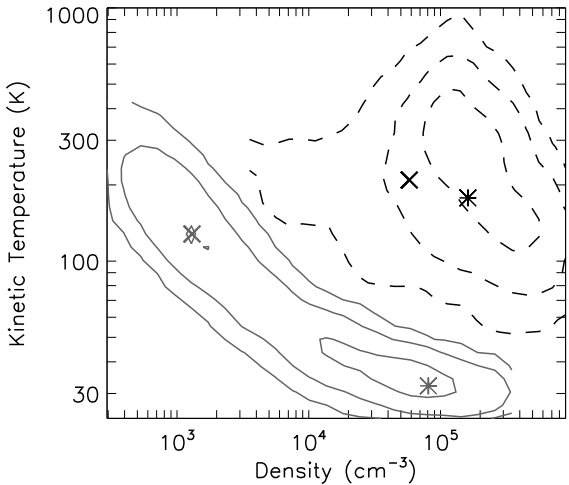}
\caption{The 2-dimensional marginalized distributions for $T_{\mathrm{kin}}$ and $n_{\mathrm{H}_2}$ for each component in the two-phase gas model described in Section~\ref{ssec:twocomp}. The contours are 68.3\%, 95.5\%, and 99.7\% assuming Gaussian distributed errors, and the stars mark the maximum likelihood solution (model ``2-Component (a)'' in Figure~\ref{fig:sled}). The crosses mark another solution that provides a good fit to the data (model ``2-Component (b)'' in Figure~\ref{fig:sled}).}
\label{fig:2comp} 
\end{center}
\end{figure}

We explore the possibility that the CO excitation for HLSW-01 arises from the combination of a dense, star-forming component and cold, lower-density gas that is not necessarily associated with star-formation. We require that the star-forming component has a gas density of $n_{\mathrm{H}_2} = 10^{3.5-6}$\,cm$^{-3}$ and a kinetic temperature of $T_{\mathrm{kin}} \ge 30$\,K, which is typical of star-forming regions in starburst nuclei \citep[e.g.][]{riechers10,carilli10,danielson10}. We then require that the other component has a lower density and temperature than the star-forming component, but a larger emitting region (or alternatively, filling fraction). We use the same prior on the velocity gradient described in Section~\ref{sec:radex} for the star-forming component, but we relax the prior on the velocity gradient of the second component so that it is not required to be self-gravitating. The 2-dimensional marginalized distributions for $T_{\mathrm{kin}}$ and $n_{\mathrm{H}_2}$ for each component are shown in Figure~\ref{fig:2comp}.

The maximum likelihood solution for this two-component model under the above constraints is $T_{\mathrm{kin}} = 180$\,K, $n_{\mathrm{H}_2} = 1.6\times10^5$\,cm$^{-3}$ for the star-forming component, and $T_{\mathrm{kin}} = 30$\,K, $n_{\mathrm{H}_2} = 8.1\times10^4$\,cm$^{-3}$ for the other component. This model is shown in Figure~\ref{fig:sled} (model ``2-Component (a)''), where the dashed red curve corresponds to the star-forming component, the dotted red curve corresponds to the cooler component, and the dot-dashed red curve shows the summed SLED from both components. The best-fit model thus consists of two dense gas phases, with warm gas distributed over a small ($330$\,pc equivalent diameter) region (assuming a filling fraction of 1), and cold gas distributed over a larger ($2.4$\,kpc diameter) region. The total molecular gas mass is $5.1\times10^{10}$\,$M_\odot$, with 30\% and 70\% of the gas mass in the warm and cold phases, respectively. The warm, dense component remains thermalized out to $J_{\mathrm{up}} = 10$, and provides the best match to our $J_{\mathrm{up}} = 9-10$ measurements, where these lines have roughly equal brightness temperatures.

However, the likelihood space for this two-component model does not have a single well defined peak, and many other solutions reproduce the lines fluxes nearly as well. For example, a warm/dense component with $T_{\mathrm{kin}} = 210$\,K, $n_{\mathrm{H}_2} = 5.8\times10^4$\,cm$^{-3}$, and a warm/moderate-density component with $T_{\mathrm{kin}} = 130$\,K, $n_{\mathrm{H}_2} = 1.3\times10^3$\,cm$^{-3}$ provides a good fit to the data as well. This model is shown by the blue curves in Figure~\ref{fig:sled} (model ``2-Component (b)''), where the dashed, dotted, and dot-dashed curves show the warm/dense and warm/moderate-density components, and their sum, respectively. In this scenario, the warm/dense gas is concentrated in a region $\sim270$\,pc, while the warm/moderate-density gas is extended over a larger region ($\sim1.4$\,kpc). The total molecular gas mass is $4.3\times10^{10}$\,$M_\odot$, with 15\% and 85\% of the gas mass in the warm/dense and warm/moderate-density component, respectively. We note that this latter component is similar to the results from our one-component fit, where the addition of a small fraction of warm/dense gas brings the $J=10\rightarrow9$ model flux in better agreement with the data.

Our data show no evidence for two-component models with a significant contribution from extended, cold, moderate-density gas with $T_{\mathrm{kin}} \sim 30-50$\,K and $n_{\mathrm{H}_2} \sim 10^{2-3.5}$\,cm$^{-3}$ like that seen in M82 \citep{panuzzo10} and many high-redshift galaxies \citep[e.g][]{riechers10,danielson10,carilli10}, as can be seen in Figure~\ref{fig:2comp}. While recent results by \citet{harris10} and \citet{ivison10b} suggest that most SMGs contain a large fraction ($\gtrsim50$\% of the total) of extended, cold gas with moderate-density, HLSW-01 appears to be an exception.

\subsection{Excitation Mechanisms}
\label{ssec:excite}

The CO measurements for HLSW-01 do not have the statistical power to discriminate between various mechanisms for heating the molecular gas, such as UV radiation from star formation in photon dominated regions (PDRs) or X-rays from an AGN in an X-ray dominated region (XDR). The $J_{\mathrm{up}} \ge 7$ line fluxes measured with Z-Spec have large uncertainties, and it is these high-$J$ transitions that can most strongly discriminate between such heating mechanisms. While high gas temperatures like that found for HLSW-01 (for both the single- and two-component models) are usually associated with heating in XDRs for sources that are known to host powerful AGNs \citep{weiss07a,ao08}, heating by cosmic rays or dissipation of turbulence can also boost the high-$J$ line fluxes; these latter mechanisms are thought to be responsible for the high excitation for M82, which does not host a strong AGN \citep{panuzzo10}. Based on the high dust temperature and 1.4\,GHz radio emission (in excess of that expected from the FIR-radio relation, C11), there is tentative evidence for a bolometrically important AGN in HLSW-01.

Obtaining higher $S/N$ measurements of the $J_{\mathrm{up}} = 7-10$ transitions -- which are all accessible from ground-based telescopes -- and spectrally resolving these lines would go a long way to help differentiate between various one- and two-phase gas models, and to better constrain the heating mechanism for the gas. In particular, high spectral resolution is needed to properly deblend the CO $J=7\rightarrow6$ line from the [\ion{C}{1}] $2\rightarrow1$ line. Directly measuring the line widths for these high-$J$ transitions can also help to determine whether they are tracing the same regions as the low-$J$ lines.

\section{Conclusions}
\label{sec:conc}

We have measured a CO redshift of $z = 2.958\pm0.007$ for the multiply-imaged, lensed galaxy HLSW-01 using Z-Spec on the CSO. Four rotational transitions of CO lie within the Z-Spec bandpass ($J_{\mathrm{up}} = 7-10$) at this redshift, and we have used a well tested algorithm (L10) to robustly identify the redshift for this source with $\gg99.99$\% confidence. This redshift has been confirmed with subsequent observations with the IRAM PdBI, CARMA, and the GBT Zpectrometer.

With six CO line measurements and one upper limit, the SLED for HLSW-01 is well sampled from low- to high-$J$, and we carry out a likelihood analysis using predictions from RADEX to constrain the molecular gas properties for this galaxy. We find that a single-component, warm/moderate-density gas with $T_{\mathrm{kin}} = 86-235$\,K and $n_{\mathrm{H}_2} = (1.1-3.5)\times10^3$\,cm$^{-3}$ provides a good fit to these data. However, several two-component gas models can describe the CO SLED as well; higher $S/N$ observations of the high-$J$ lines with finer spectral resolution are needed in order to differentiate between these scenarios.

Based on the nearly equal line luminosities measured for the $J=1\rightarrow0$, and $J=3\rightarrow2$ transitions, these data are inconsistent with models that require an extended, cold/moderate-density gas component. This is in contrast to recent results that show that the CO $J=1\rightarrow0$ line flux for most high-redshift SMGs, as well as nearby starburst galaxies such as M82, requires the existence of such a low-excitation component. High spatial resolution CO maps of both low- and high-$J$ lines with CARMA, the IRAM PdBI, and/or the EVLA could potentially provide more concrete information on the presence and distribution of cold and warm molecular gas within this galaxy.

\acknowledgments

We thank the anonymous referee for their suggestions, which improved the presentation of these results. We also thank the staff at the CSO for support of the Z-Spec observations, and we appreciate the help of Robert Hanni and Jon Rodriguez with observing. JEA acknowledges support by NSF grant AST-0807990 and by the CSO NSF Cooperative Agreement AST-0838261. JK acknowledges funding through an NSF graduate research fellowship. AJB acknowledges support from NSF grant AST-0708653. Z-spec was constructed under NASA SARA grants NAGS-11911 and NAGS-12788 and an NSF Career grant (AST-0239270) and a Research Corporation Award (RI0928) to JG, in collaboration with the Jet Propulsion Laboratory, California Institute of Technology, under a contract with the National Aeronautics and Space Administration. We acknowledge Peter Ade and his group for their filters and Lionel Duband for the 3He / 4He refrigerator in Z-Spec, and are grateful for their help in the early integration of the instrument.

%% To help institutions obtain information on the effectiveness of their
%% telescopes, the AAS Journals has created a group of keywords for telescope
%% facilities. A common set of keywords will make these types of searches
%% significantly easier and more accurate. In addition, they will also be
%% useful in linking papers together which utilize the same telescopes
%% within the framework of the National Virtual Observatory.
%% See the AASTeX Web site at http://www.journals.uchicago.edu/AAS/AASTeX
%% for information on obtaining the facility keywords.

%% After the acknowledgments section, use the following syntax and the
%% \facility{} macro to list the keywords of facilities used in the research
%% for the paper.  Each keyword will be checked against the master list during
%% copy editing.  Individual instruments or configurations can be provided 
%% in parentheses, after the keyword, but they will not be verified.

{\it Facilities:} \facility{CSO (Z-Spec)}, \facility{GBT (Zpectrometer)}, \facility{IRAM (PdBI)}, \facility{CARMA}, \facility{SMA}, \facility{{\it Herschel} (SPIRE)}

\bibliographystyle{apj}
\bibliography{KScott_HLSW01_ZSpec_refs}

\end{document}